\documentclass[11pt,twoside]{article}

\usepackage{asp2006}
\usepackage{lscape}

\usepackage{graphicx}
\usepackage{sidecap}
\usepackage{adassconf}

\begin{document}
\title{The Vertical Dust Structure in Spiral Disks}
\author{B. W. Holwerda\altaffilmark{1}, P. Kamphuis\altaffilmark{2},  R. J. Allen\altaffilmark{1}, R. F. Peletier\altaffilmark{2}, and P. C. van der Kruit\altaffilmark{2}}
\altaffiltext{1}{Space Telescope Science Institute} 
\altaffiltext{2}{Kapteyn Astronomical Institute}

\begin{abstract} 
The halo of NGC 891 has been the subject of studies for more than a decade. One of its most striking features is the large asymmetry in H$\alpha$  emission. We have taken a quantitative look at this asymmetry at different wavelengths for the first time. We propose that NGC 891 is intrinsically almost symmetric, as seen in {\em Spitzer} observations, and the large asymmetry in H$\alpha$ emission is mostly due to dust attenuation. We quantify the additional optical depth needed to cause the observed H$\alpha$ asymmetry. A comparison of large strips on the North East side of the galaxy with strips covering the same area in the South West we can quantify and analyze the asymmetry in the different wavelengths. From the 24 $\mu$m emission we find that the intrinsic asymmetry in star-formation in NGC 891 is small i.e., approximately 30\%. The additional asymmetry in H$\alpha$ is modeled as additional symmetric dust attenuation which extends up to $\sim$ 40'' (1.9 kpc) above the plane of the galaxy with a mid-plane value of  $\tau$=0.8 and a scale height of 0.5 kpc. This observational technique offers the possibility  to quantify the effects of vertical ISM disk stability as an explanation for dust lanes in massive galaxies \citep{Dalcanton04}.
\end{abstract}





\section{Dust in and out of the disk}

Extinction by dust manifests itself most conspicuously in the dark dust lanes seen in optical images of edge-on galaxies. Occasionally dark structures can be observed extending out of the plane \citep{Howk97, Howk99a, Howk99b, Thompson04, Howk05}. The study of these dust-rich structures is hampered by the need to have enough stellar light backlighting the dark structures. 
Previous observations of the scale-height in nearby, mostly edge-on, galaxies appeared to find small scale-heights for dust \citep[e.g,][]{Xilouris99,Holwerda05b,Bianchi07} but recent observations indicate a much higher scale-height for the dust \citep[e.g.,][]{Seth05}, one that is similar to the stellar scale-height.

\begin{figure}[h]
  \centering  
     \includegraphics[width=\textwidth]{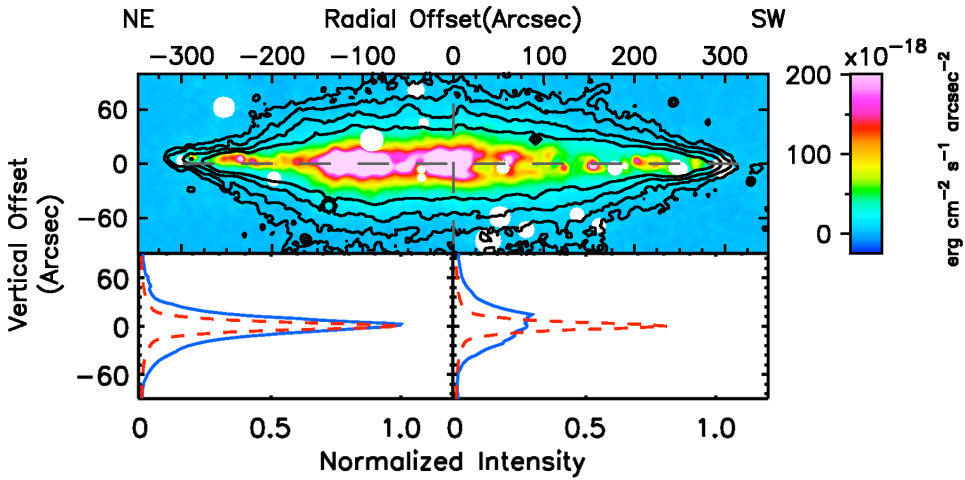}\\
\caption{\label{f:map} The H$\alpha$ emission \citep[false color from][]{Rand94} with 24 $\mu$m contours overlaid, below are the integrated vertical profiles in H$\alpha$(blue) and 24 $\mu$m (red dashed line), normalized to the top in the NE side. Both types of emission are powered by ongoing star-formation. The 24 $\mu$m is very symmetrical while the H$\alpha$ is not.  Figure from \cite{Kamphuis07}.}
\end{figure}

In many edge-on galaxies an H$\alpha$ halo can be observed. \cite{Heald06} and \cite{Kamphuis07a} reported on the kinematics of this H$\alpha$ halo for NGC 891. Its H$\alpha$ distribution is extremely asymmetric: the NE side halo is much more extended and brighter (Figure \ref{f:map}). Is this due to an intrinsic asymmetry in the H$\alpha$ emission or an extinction effect? 

\begin{figure}[h]
  \centering  
       \includegraphics[width=0.9\textwidth]{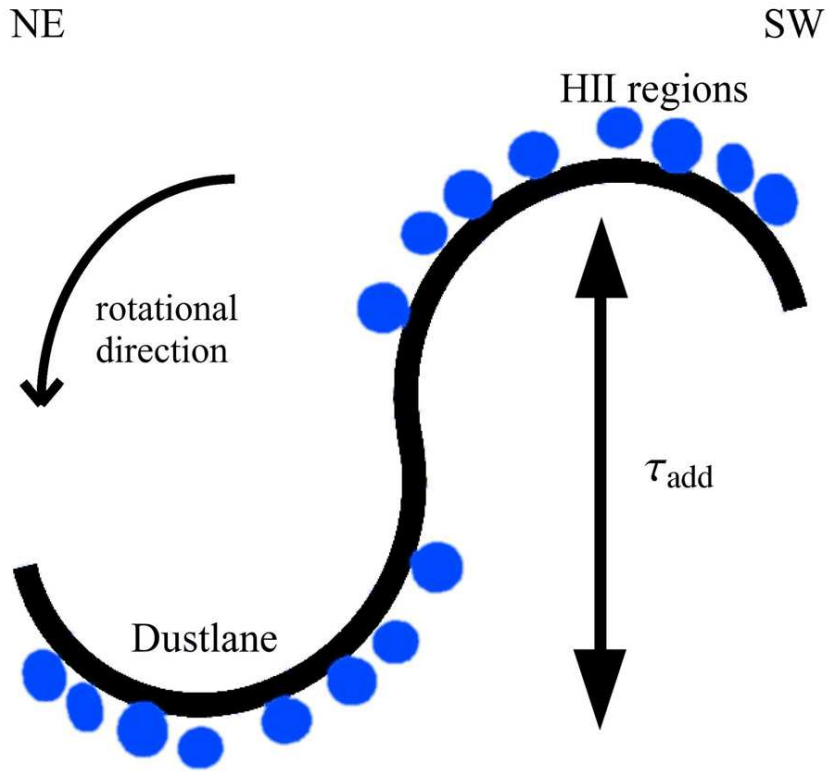}\\
\caption{\label{f:mod} A sketch of our model of NGC891. The emission originates predominantly from HII regions (blue) and much of the dust is in the spiral arms (thick black S-shape). The H$\alpha$ emission is more absorbed in the SW case ($\tau_{add}$).  Figure from \cite{Kamphuis07}. }
\end{figure}

\begin{figure}[h]
  \centering  
       \includegraphics[width=\textwidth]{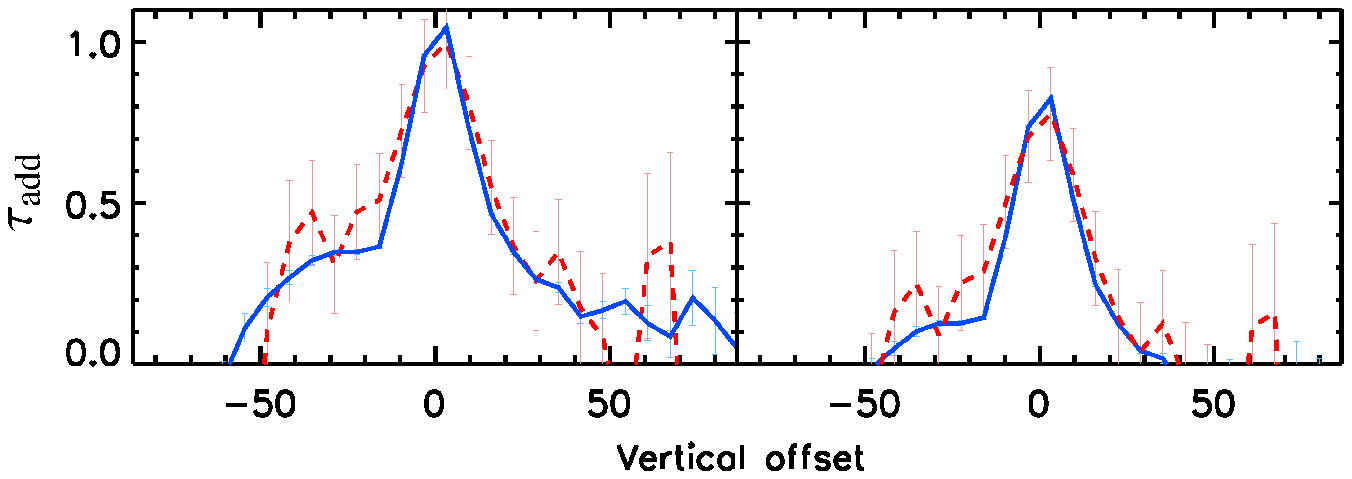}
\caption{\label{f:prof} The vertical extent in arseconds of the additional absorption ($\tau_{add}$), assuming perfect NE/SW symmetry (left) and taking the 24 $\mu$m symmetry as the intrinsic one (right). The blue solid line is from the \cite{Rand94} H$\alpha$ data and the dashed red line based on the \cite{Kamphuis07a} H$\alpha$ data. Figure from \cite{Kamphuis07}.}
\end{figure}

In \cite{Kamphuis07}, we compared two star-formation driven emissions (H$\alpha$ and {Spitzer} 24 $\mu$m). The 24 $\mu$m is much more symmetric (Figure \ref{f:map}). Assuming both types of emission originate from star-formation regions in the spiral arms of NGC 891 and that the NE arm is closest to us (Figure \ref{f:mod}), the difference in symmetry between the two types of emission can be attributed to extinction in the inner disk extending out of the plane of the disk ($\tau_{add}$). The inferred scale-height of the out-of-the plane dust (Figure \ref{f:prof}) is similar to the heights to which \cite{Howk97} report dark structures in NGC 891. Still this height is much less than the height of dust emission in the halo as reported by M. Burgdorf \& M. Ashby, ({\em this volume}). 

 \cite{Dalcanton04} link the appearance of dust lanes to the vertical stability of spiral disks. In more massive disks, gravitation wins over the star-formation driven turbulence and the dust-rich ISM collapses into the characteristic dust lane, notably in the inner scale-length of the disk. In contrast, low-mass galaxies display a more flocculant appearance of dust structures. The additional extinction out-of-the plane of NGC 891, a massive galaxy seems to be in contradiction to this. However, NGC 891 is undergoing significant star-formation ($log( ~ L(FIR)~) ~ = ~ 10.42 ~ L_\odot $) and much of its ISM is in the dustlane.

\section{Future work}

H$\alpha$ haloes are relatively common \citep{Rossa03} around edge-on spiral galaxies. {\em Spitzer}/MIPS observations are available for 28 edge-ons. Many of these show signs of asymmetric H$\alpha$ emission in and out of the plane of the disk. This offers the opportunity to analyze a sample of edge-ons in the same way as explored in \cite{Kamphuis07}. 

The symmetry analysis can be extended to include lower-mass disks and disks with different levels of star-formation. Both are an input into the vertical disk stability model of \cite{Dalcanton04}. The advantages of the method are (1) a measure of the ISM's vertical extent right around the inner part of the spiral disks and (2) the measurement is independent of the stellar distribution, often needed to backlight or heat the dust in other measurements of dust thickness. An additional benefit is that the method can be applied to a data-set with a widely inhomogeneous data-quality because we compare the left and right symmetries of the images. {\it Spitzer} 24 $\mu$m observations resolve the vertical structure of some galaxies but the resolution is not critical for the method because the {\it Spitzer} observations are used to verify the symmetry argument for the H$\alpha$ data.

The drawbacks of the method are: (1) the assumption of a direct spatial relation between the H$\alpha$ and 24 $\mu$m emission. This may well not hold true in low-level starformation regions or well outside the disk. And (2), the need for a specific spiral arm geometry. The position of the spiral arms in NGC 891, is undoubtedly the most favorable. The method may not work as well if the spiral arms are positioned different than in Figure \ref{f:mod} or if the spiral pattern is flocculant instead of grand-design.

The physical conditions in spiral disks determine the ISM morphology: the turbulence and energy input by star-formation versus the local gravitational pull by the disk's mass in gas and stars. What are the global conditions in a disk for a dustlane to form? Symmetry measurements of edge-on spiral disks in wavelengths that are strongly affected by extinction --H$\alpha$-- and those that are not --24 $\mu$m-- can shed light on the global conditions that prevail in disks when dust lanes form.





\end{document}